\documentclass[useAMS,usenatbib]{mn2e}

\usepackage{times}
\usepackage{bm}
\usepackage{graphicx}
\usepackage{dcolumn}%
\usepackage{subfigure}

\begin{document}
\journal{astro-ph/0512484}

\title[Model selection and dark energy surveys]
{Model selection as a science driver for dark energy surveys}
\author[P. Mukherjee {\it et al.}]
{Pia Mukherjee,$^1$ David Parkinson,$^1$ Pier Stefano Corasaniti,$^2$
Andrew R. Liddle$^1$\cr and Martin Kunz$^3$\\ 
\vspace*{-6pt} {\small \em $^1$Astronomy Centre, University of Sussex,
Brighton BN1 9QH, United Kingdom}\\
\vspace*{-6pt} {\small \em $^2$ISCAP, Columbia University, New York,
NY 10027, USA}\\
\vspace*{-6pt} {\small \em $^3$Department of Theoretical Physics,
University of Geneva, 24 quai Ernest Ansermet, CH-1211 Geneva 4,
Switzerland}}
\maketitle

\begin{abstract}
A key science goal of upcoming dark energy surveys is to seek time
evolution of the dark energy. This problem is one of {\em model
selection}, where the aim is to differentiate between cosmological
models with different numbers of parameters.  However, the power of
these surveys is traditionally assessed by estimating their ability to
constrain parameters, which is a different statistical problem.  In
this paper we use Bayesian model selection techniques, specifically
forecasting of the Bayes factors, to compare the abilities of
different proposed surveys in discovering dark energy evolution. We
consider six experiments --- supernova luminosity measurements by the
Supernova Legacy Survey, SNAP, JEDI, and ALPACA, and baryon acoustic
oscillation measurements by WFMOS and JEDI --- and use Bayes factor
plots to compare their statistical constraining power.  The concept of
Bayes factor forecasting has much broader applicability than dark
energy surveys.
\end{abstract}


\section{Introduction}

Uncovering the nature of dark energy in the Universe is perhaps the
greatest challenge facing cosmologists in coming years. In recent
months many proposed experiments to probe dark energy have been
defined, especially in response to a call for white papers by the Dark
Energy Task Force set up jointly in the US by the NSF, NASA and
DOE. These propose a variety of techniques to constrain dark energy
parameters, including the luminosity distance--redshift relation of
Type Ia supernovae (SNe-Ia), the angular-diameter distance--redshift
and expansion rate--redshift relations measured by baryon acoustic
oscillations, and use of weak gravitational lensing to probe the
growth rate of structures.

Following on from heritage of CMB anisotropy studies, the standard
tool used to illustrate the power of a given instrument or survey is a
plot of the projected parameter errors around one or more fiducial
models, estimated using a Fisher information matrix approach or
likelihood analysis of Monte Carlo simulated data (Knox 1995; Jungman
et al.~1996; Zaldarriaga, Spergel \& Seljak 1997; Bond, Efstathiou \&
Tegmark 1997; Efstathiou \& Bond 1999).  Typically, a projection of
the parameter uncertainties onto a two-parameter equation-of-state
model for dark energy is deployed, showing how tightly parameters are
expected to be constrained around, for instance, the cosmological
constant model. The implication is intended to be that if the true
values lie outside those error ellipses, then the survey will be able
to exclude the cosmological constant model.

However, the principal goal of such surveys is usually identified as
being the discovery of dark energy evolution. This is not a parameter
estimation question, but rather one of {\em model selection} (Jeffreys
1961; MacKay 2003; Gregory 2005), where one seeks to compare
cosmological models with different numbers of variable
parameters. Within the framework of Bayesian inference, the
statistical machinery to make such comparisons exists, and is based
around statistics known as the Bayesian evidence and the Bayes
factor. The Bayes factor has the literal interpretation of measuring
the change in relative probabilities of two models in light of
observational data, updating the prior relative model probabilities to
the posterior relative model probabilities.

In this paper we use Bayesian model selection tools to assess the
power of different proposed experiments. Our method is related to the
Expected Posterior Odds (ExPO) forecasting recently developed by
Trotta (2005).  The main difference is that he takes the present
observational constraints on the extended model, and seeks to estimate
the fraction of that parameter space within which that model can be
distinguished from a simpler embedded model. By contrast, we take a
theoretically-motivated view of the parameter space of interest, and
seek the locations within that parameter space corresponding to dark
energy models which are distinguishable from a cosmological constant
by a given experiment.  We also differ computationally, in that as
well as using approximate techniques, we use the nested sampling
algorithm of Skilling (2004), as implemented by Mukherjee, Parkinson
\& Liddle (2006), to compute the evidences accurately numerically.

The paper is organized as follows. In Section~\ref{bsf} we introduce
model selection in the Bayesian framework.  Section~\ref{surveys}
describes the dark energy surveys we make model selection forecasts
for, and Section~\ref{results} presents the results.  We conclude in
Section~\ref{Conclusions}. We consider some additional technical
details and review the standard parameter forecast procedure in two
Appendices.

\section{Bayesian model selection}\label{bsf}

\subsection{The model selection framework}

The Bayesian model selection framework has now been described in a
variety of places (Jaffe 1996; MacKay 2003; Marshall, Hobson \& Slosar
2003; Saini, Weller \& Bridle 2004; Gregory 2005; Trotta 2005;
Mukherjee et al.~2006) and we will keep our account brief.

In this context, a model is a choice of parameters to be varied to fit
the data, its predictions being reflected in the prior ranges for
those parameters. A model selection statistic aims to set up a tension
between model complexity and goodness of fit to the observed data,
ultimately providing a ranked list of models based on their
probabilities in light of data. Within Bayesian inference, the
appropriate statistic is the {\em Bayesian evidence} $E$ (also known
as the marginal likelihood), which is the probability of the data
given the model in question.  It is given by integrating the
likelihood $P(D|\theta,M)$ over the set of parameters $\theta$
of model $M$, in light of data $D$, i.e.
\begin{equation}
E(M) \equiv P(D|M)=\int d\theta \, P(D|\theta,M)P(\theta|M)\,,
\label{evdef}
\end{equation}
where the prior $P(\theta|M)$ is normalized to unity. The evidence is
thus the average likelihood of the model over its prior parameter
space. Rather than focusing simply on the best-fit parameters (which
will always tend to favour the most complex model available), it
additionally rewards models with good predictiveness.

By Bayes Theorem the evidence updates the prior model probability to
the posterior model probability. The ratio of the evidences of two
models, $M_0$ and $M_1$, is known as the Bayes factor (Kass \& Raftery
1995):
\begin{equation}
B_{01} \equiv \frac{E(M_0)}{E(M_1)} \,.
\end{equation}
Note that the prior model probabilities are to be chosen in the
Bayesian approach, and different people may have different opinions as
to those.  Nevertheless, everyone will agree on whether the Bayes
factor led to their original belief becoming more or less tenable
relative to another model in light of the data.  In describing results
from Bayes factors, it is common to presume that the prior model
probabilities are equal, and we shall follow that practice; anyone who
thinks otherwise can readily recalculate the posterior relative model
probability.

The Bayesian evidence provides a ranked list of the models in terms of
their probabilities, obviating the need to specify an arbitrary
significance level as in frequentist chi-squared tests. Nevertheless
one still has to decide how big a difference will be regarded as
significant.  A useful guide as to what constitutes a significant
difference between models is given by the Jeffreys' scale (Jeffreys
1961); labelling as $M_0$ the model with the higher evidence, it rates
$\ln B_{01} < 1$ as `not worth more than a bare mention', $1<\ln
B_{01} < 2.5$ as `substantial', $2.5< \ln B_{01} < 5$ `strong' to
`very strong' and $5<\ln B_{01}$ as `decisive'. Note that $\ln
B_{01}=5$ corresponds to odds of 1 in about 150, and $\ln B_{01}=2.5$
to odds of 1 in 13.

\subsection{Forecasts and the Bayes factor plot}

\label{ss:list}

In order to forecast the power of an experiment for model selection,
we ask the following question: Given a well-motivated simpler model
embedded in a larger parameter space, how far away does the true model
have to lie in order that the experiment is able to exclude the
simpler model?  There are many such cases present in cosmology, for
example $\Lambda$CDM in the space of evolving dark energy models, the
question of whether we live in a spatially-flat universe, or whether
the initial power spectrum of perturbations is exactly scale
invariant, or exactly a power law, etc. Here we will use the dark
energy as a worked example.  The Bayesian evidence of models with dark
energy has been computed from current observational datasets by
several authors (Saini et al.~2004; Bassett, Corasaniti \& Kunz 2004;
Mukherjee et al.~2006), all finding that the simple $\Lambda$CDM model
is the preferred fit to present data. Our aim here is to forecast its
outcome in light of future datasets, in order to assess the power of
those surveys for model selection.

Our procedure is as follows. We first select an experimental
configuration.  We then consider a set of `fiducial
models' characterized by parameter values $\hat{\theta}$, which we
shall consider in turn to be the true model.  For each choice of
fiducial model in our dark energy space we generate a set of simulated
data $D$ with the properties expected of that experiment. We then
compute the evidences of the two models we seek to distinguish, here
the $\Lambda$CDM model and the general dark energy model.  For
definiteness, we choose to assess a set of dark energy experiments by
their ability to distinguish a $\Lambda$CDM model from a two-parameter
dark energy model with equation of state given by
\begin{equation}
w(z)=w_0+w_a(1-a),\label{lp} 
\end{equation}
where $w_0$ and $w_a$ are constants and $a$ is the scale
factor. Although the latter is sometimes referred to as the Linder
parametrization based on its use in Linder (2003), it appears to have
been first introduced by Chevallier \& Polarski (2001).

Here $\hat{\theta}$ refers to all the parameters of the model, but we
are principally interested in the dependence of the Bayes factor on
the extra parameters characterizing the extended model, here $w_0$ and
$w_a$. Our main plots therefore show the difference in log evidence
between the $\Lambda$CDM model and the two-parameter evolving dark
energy model, plotted in the $w_0$--$w_a$ plane.  This is the Bayes
factor plot, which is presented in Section \ref{results} for different
dark energy surveys, with contours showing different levels at which
the two models can be distinguished by data simulated for each
experiment.

In general the Bayes factor is a function of all the fiducial
parameters, not just the dark energy ones. For the dark energy
application this dependence turns out to be unimportant, but for
completeness we discuss some issues relating to this in Appendix~A.

Use of the Bayes factor plots to quantify experimental capabilities is
quite distinct, both philosophically and operationally, from the use
of parameter error forecasts; for readers unfamiliar with the latter
we provide a short review in Appendix~B. We highlight the advantages
of the Bayes factor approach as follows:
\begin{enumerate}

\item Most experiments, particularly dark energy experiments, are
motivated principally by model selection questions, e.g.~does the dark
energy density evolve, and so should be quantified by their ability to
answer such questions.

\item In Bayes factor plots, the data are simulated at each point of
the dark energy parameter space that is to be confronted with the
simpler $\Lambda$CDM model, whereas parameter error forecasts are
plotted around only selected fiducial models (often just one). In
particular, in the latter case the data are usually simulated for a
model that people hope to exclude, rather than the true model which
would allow that exclusion.

\item The Bayesian model selection procedure accords special status to
the $\Lambda$CDM model as being a well-motivated lower-dimensional
model, which in Bayesian terms is rewarded for its predictiveness in
having a smaller prior volume.  Parameter estimation analyses do not
recognize a special status for such models, e.g.~the same criterion
would be used to exclude $w=-0.948$ as $w=-1$. Model selection
criteria provide a more stringent condition for acceptance of new
cosmological parameters than parameter estimation analyses. Model
selection analyses can also accrue positive support for the simpler
model, whereas parameter estimation methods can only conclude
consistency of the simpler model.

\item In parameter error studies, it is necessary that the simple
model is embedded as a special case of the second model. While the
models we discuss here are indeed of that type, the Bayes factor could
also be used to compare non-nested models (e.g.~two different types of
isocurvature perturbation).

\item Although it is not essential to do so, most parameter estimation
forecasts assume a Gaussian likelihood in parameter space, while the
Bayes factor plot uses the full likelihood.

\end{enumerate}

Set against these advantages, the only disadvantages of the Bayes
factor method are that it is computationally more demanding, and that
its conceptual framework has yet to become as familiar as that of
parameter estimation.

\subsection{Bayes Factor Evaluation}\label{method}

We use the nested sampling algorithm (Skilling 2004; Mukherjee et
al.~2006), which is fast enough to enable exact evaluations of the
evidence for many fiducial parameter values. For comparison we also
compute results with the Savage--Dickey method outlined in Trotta
(2005), using a Fisher matrix approximation to the likelihood about
the true model, given as equation~(\ref{fish}) in Appendix~B. We
discuss how the results from the two methods compare in one case, and
present our main results using the more accurate nested sampling
method.

\subsubsection{Nested Sampling Algorithm}

The Bayes factor can be found by calculating the evidences of the two
models independently, and then taking their ratio. This method
requires integration over the extra cosmological parameters, which
does not feature in the Savage--Dickey method. Here we use our
implementation of the nested sampling algorithm (as described in
Mukherjee et al.~2006) to perform the integration.  To quickly
summarize, the algorithm (Skilling 2004) recasts the problem as a
one-dimensional integral in terms of the remaining `prior mass' $X$,
where $dX = P(\theta|M)d\theta$.  The integral becomes
\begin{equation}
E = \int_0^1 L(X) dX \,,
\end{equation}
where $L(X)$ is the likelihood $P(D|\theta,M)$.  The algorithm samples
the prior a large number of times, assigning an equal prior mass to
each sample.  The samples are then ordered by likelihood, and the
integration follows as the sum of the sequence,
\begin{equation}
E = \sum_{j=1}^m E_j\,, \quad E_j=\frac{L_j}{2}(X_{j-1}-X_{j+1}) \,,
\end{equation}
where the lowest likelihood sample goes into the sum, and is discarded
to be replaced by a new sample selected under the condition that it
lies above the likelihood of the discarded sample. In this way the
algorithm works its way in to the highest likelihood peak.

We compute the evidences using 300 live points, averaging over six
repetitions of the calculation. This requires approximately $10^4$
likelihood evaluations per evidence computation.

\subsubsection{Savage--Dickey Formula} 

Bayes factors for two nested models can be computed using the
Savage--Dickey density ratio (Dickey 1971; Verdinelli \& Wasserman
1995; see Trotta 2005 for an application to cosmological model
selection).  Assuming a Gaussian approximation to the likelihood, the
Savage--Dickey formula of an extended model $M_1$ with two free model
parameters ($\hat{\theta}_1$,$\hat{\theta}_2$) and flat priors
($\Delta\theta_1,\Delta\theta_2$), versus a simpler model $M_0$ with
$\hat{\theta}_1=\hat{\theta}^{*}_1$ and
$\hat{\theta}_2=\hat{\theta}^{*}_2$, is
\begin{equation}
B_{01}(\hat{\theta}_1,\hat{\theta}_2)=\frac{\Delta\theta_1
\Delta\theta_2} {2\pi\sqrt{\det{F^{-1}}}} 
e^{-\frac{1}{2}\sum_{\mu\nu}(\hat{\theta}_\mu-\hat{\theta}^{*}_\mu)
F_{\mu\nu} (\hat{\theta}_\nu-\hat{\theta}^{*}_\nu)}\,,\label{savage}
\end{equation}
where $F_{\mu\nu}$ is the marginalized $2\times2$ Fisher matrix
evaluated at $\hat{\theta}_\mu$.  Our conventions are defined in
Appendix~B, and we have used the hat sign for the extended model
parameters to emphasize that the Bayes factors directly compare the
fiducial models of the parameter estimation analysis to the simpler
nested model.
 
In our specific case, $M_1$ consists of all dark energy models
parametrized by different values of $w_0$ and $w_a$, while $M_0$ is
the cosmological constant model which is nested in $M_1$ with $w_0=-1$
and $w_a=0$.  We use equation~(\ref{savage}) to compute the
Bayes factor as function of $w_0$ and $w_a$, to determine the range of
dark energy models that a given experiment is able to distinguish from
$\Lambda$CDM.

{}From equation~(\ref{savage}) we can see that the Bayes factor depends on
two multiplicative terms, namely an exponential factor and an overall
amplitude. The former accounts for the distance in the parameter space
of the model $M_1$ from $M_0$ in units of the forecasted parameter
uncertainty.  The latter accounts for the fraction of the accessible
prior volume of the extended model $M_1$ in light of the data, and
hence this factor penalizes the model $M_1$ for having a large
parameter space compared to model $M_0$.  As shown in Trotta (2005),
this factor can be interpreted as an estimate of the informative
content of the data,
\begin{equation}
I=\log_{10}\frac{\Delta\theta_1\Delta\theta_2}{\sqrt{\det{F^{-1}}}} \,,
\end{equation}
being the order of magnitude by which the prior volume of model $M_1$
will be reduced by the arrival of the forecasted data.

\section{Dark energy surveys}\label{surveys}

\subsection{The surveys}

We have simulated observational data for two types of future dark
energy experiments: luminosity distance probes made through the
measurement of Type Ia supernovae, and angular-diameter
distance measurements from baryonic acoustic oscillations (BAO). Some
of the experiments considered have weak lensing parts too (SNAP, JEDI,
ALPACA), but we do not derive dark energy constraints from simulated
weak lensing measurements here.\footnote{Both SN-Ia and BAO are
distance indicators, while weak lensing is sensitive to growth and
dark energy perturbations. Complementarity of weak lensing with
SN-Ia/BAO will thus be very interesting in probing dark energy more
comprehensively.} Note that all these experiments are presently
undergoing optimization of their survey structure which may improve
their science return.

For the SNe-Ia, we compared four different surveys. The CFHT SuperNovae
Legacy Survey (SNLS) is already underway but we consider the full
five-year survey, while the SuperNovae Acceleration Probe (SNAP) and
the Joint Efficient Dark Energy Investigation (JEDI) satellite
missions, plus the Advanced Liquid-mirror Probe for Astrophysics,
Cosmology and Asteroids (ALPACA) ground-based survey, are all proposed
experiments. For all experiments we assumed the same spread in
magnitude $\delta_m=0.18$ of the supernovae, representing the combined
effect of measurement error and intrinsic dispersion in the
light-curve corrected luminosity [the intrinsic dispersion alone was
recently estimated as 0.12 mag by SNLS (Astier et al.~2006)].  We used
the number distribution of SNe-Ia with redshift for the different
surveys as outlined in the literature; the total numbers used are 700,
2\,000, 13\,000 and 86\,000 for SNLS (Pritchet et al.~2004; Astier et
al.~2006), SNAP (Aldering et al.~2004), JEDI (Crotts et al.~2005) and
ALPACA (Corasaniti et al.~2005) respectively.  We also assumed all
surveys would have support from an extra 300 nearby SNe-Ia observed by
ground-based telescopes in the redshift range $0.03< z <0.08$, which
also had a slightly smaller spread in magnitude ($\delta_m=0.15$). We
assumed no systematic errors in any of the magnitudes, only
statistical errors (except for one comparison case shown later).

For the baryonic acoustic oscillations we compared two different
surveys, the ground-based Wide-Field Fibre-fed Multi-Object
Spectrograph (WFMOS) and the satellite mission JEDI (JEDI will perform
both a SN-Ia survey and a BAO survey). The BAO surveys measure both
angular-diameter distance $D_{{\rm A}}(z)$ and the Hubble parameter
$H(z)$ in a series of redshift bins. We calculated the expected errors
of the measurements in each bin using the Fisher matrix approach of
Seo \& Eisenstein (2003), marginalizing over the physical matter
density $\Omega_{{\rm m}} h^2$.

In order to obtain accurate results from experiments of these types,
it is necessary that strong degeneracies with the matter density are
removed by bringing in constraints from other sources. We make the
assumption that by the time these surveys are operative, data
compilations including Planck satellite observations will have
provided a measurement of $\Omega_{{\rm m}}$ to an accuracy of $\pm
0.01$ (see for example Pogosian et al.~2005). We include such a
measurement by adding an extra term to the likelihood centred around
the fiducial density parameter value. In the absence of such external
information, dark energy surveys would give a much poorer return. We
will briefly explore the effect of varying this assumption in
Section~\ref{s:variations}.  Similarly in the BAO case we assume a 1\%
measurement uncertainty on $\Omega_{{\rm m}}h^2$ (see for example
Tegmark et al.~2000).

\begin{figure*}
\centering
\includegraphics[width=8cm]{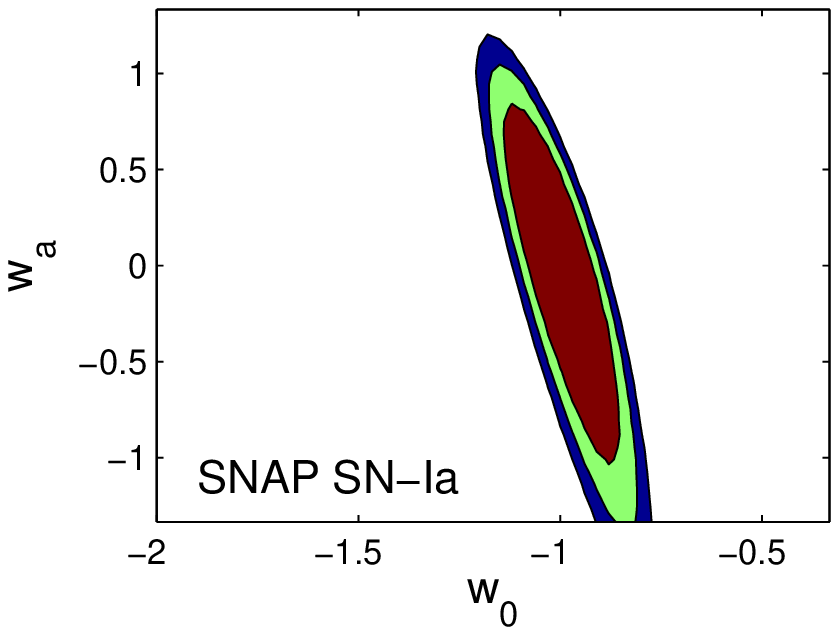}
\hspace*{1.2cm}
\includegraphics[width=8cm]{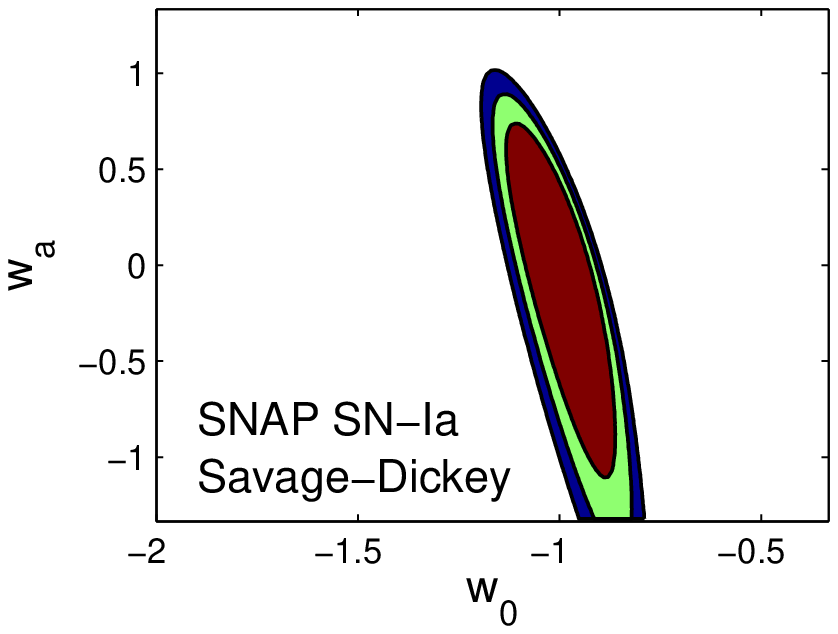}
\caption{The Bayes factor plot for the SNAP mission supernova
survey. The left panel shows the calculation using nested samping, and
the right plot using the Savage--Dickey formula with the Fisher
information matrix. The contour levels are $\ln B_{01}$ equal to $0$,
$-2.5$ and $-5$.}
\label{fig:SNAP}
\end{figure*}

\subsection{Priors}

The model priors are the parameter ranges over which the evidence
integral is carried out.  Ordinarily in model selection these are
supposed to be the wide priors seen as appropriate when the model was
first considered, and not those motivated by current data. If one
allows the model priors to `follow the data' into a small region of
parameter space, then model selection calculations will always be
inconclusive in the long term, as this requires each new experiment to
exclude a model again on its own, rather than the cumulative effect of
all observations.\footnote{An alternative, equivalent, view more in
the Bayesian spirit is that one can update the model prior ranges
after new data, provided one also updates the model probabilities and
keeps track of them as well. In practice, cosmological data analysis
tends to re-apply a broad set of data to models with wide priors each
time, which is consistent with the model selection philosophy.} The
precise results for the Bayes factor will have some dependence on the
choice of priors (see below), though the effect of the choices on
model comparison or on survey comparison is diminished as the same
priors are used for the common model parameters and the same priors
are used for each survey being compared.

Our choices are as follows. We only consider flat Universes, so that
$\Omega_\Lambda=1-\Omega_{{\rm m}}$.  For the model priors, we impose
the prior ranges $-2 < w_0 < -0.333$ and $-1.333 < w_a < 1.333$ on the
interesting parameters, and $0<\Omega_{{\rm m}}<1$ and $0.5 < h < 0.9$
on the other parameters (the Hubble parameter is needed only for the
baryon oscillation probes). The fiducial values for $\Omega_{{\rm m}}$
and $h$ are taken to be 0.27 and 0.7 respectively.

Note that for the phenomenological two-parameter evolving dark energy
model, the model priors on $w_0$ and $w_a$ that we have chosen to work
with are somewhat arbitrary. However if the prior space were reduced
for instance by a factor of 2, that would increase the $\ln E$ of the
evolving dark energy model by at most $\ln 2 \simeq 0.69$, and this
would not significantly affect our contours or conclusions which are
based on differences in $\ln E$ of 2.5 and 5.  We make a brief
investigation of some prior dependences in Section~\ref{s:variations}.

One should note that our conclusions also depend to some extent on our
chosen dark energy parametrization being able to describe the true
model. One could consider more general cases, such as the
four-parameter models of Corasaniti \& Copeland (2003) and Linder \&
Huterer (2005).  For the purpose of assessing the power of an
experimental proposal, it seems reasonable to presume that experiments
capable of distinguishing two-parameter models are likely also to be
better under other parametrizations. If the effect of dark energy were
in fact a non-smooth variation in the equation of state and a
non-smooth variation of the expansion history with redshift, then our
results are too optimistic; the validity of reparametrizing the
observables, which are the expansion history in different redshift
bins as measured by the surveys, into ($w_0$,$w_a$) would need to be
tested when the data arrive. Aspects of parametrization have been
explored in Wang \& Tegmark (2004) and Bassett et al.~(2004).

\section{Results}\label{results}

\subsection{Comparison of calculational techniques}

We begin by comparing our two methods of computing the Bayes factor,
focussing on the SNAP mission supernova survey. The Bayes factor plots
are shown in Figure~\ref{fig:SNAP}.  In the left panel we plot
isocontours of Bayes factors in the $\hat{w}_0$--$\hat{w}_a$ plane
inferred from the nested sampling method.  The plot shows the generic
structure expected of Bayes factor plots.  In the central region, the
simulated data are for models very close to $\Lambda$CDM, so that
model gives a good fit and is further rewarded for its predictiveness,
giving a positive Bayes factor which would support $\Lambda$CDM over
the dark energy model.  At the zero contour (the innermost one
plotted) the models fare equally well, and then at greater distances
the dark energy model becomes favoured.  If the true parameters lie
outside those contours, SNAP will be able to exclude $\Lambda$CDM at
the probability corresponding to the contour level.

We see a strong degeneracy between the two parameters, meaning that
supernova data are not good at constraining models in this particular
parameter direction. This same degeneracy shows up in the usual Fisher
matrix error projection method. Its precise direction depends on the
redshift distribution of the supernovae.

\begin{figure}
\centering
\includegraphics[width=8cm]{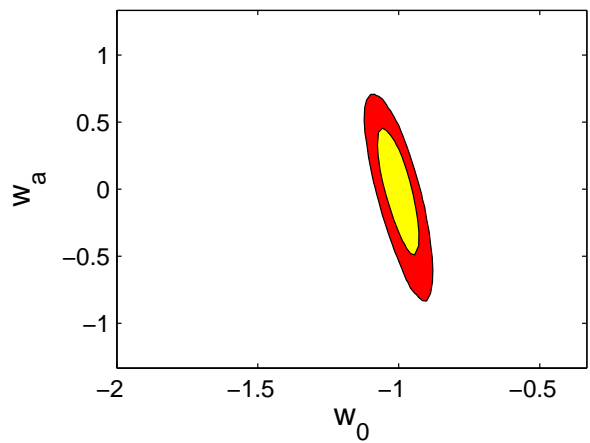}
\caption{This plot shows a parameter error forecast for the SNAP SN-Ia
experiment, taking $\Lambda$CDM as the true model. The contour levels
indicate 68\% and 95\%. While this figure uses the full likelihood, in
this case a Gaussian approximation using the Fisher matrix gives
essentially identical results.}
\label{fig:SNAP2}
\end{figure}

In the right panel we plot the projected Bayes factor contours derived
from the Savage--Dickey formula for the same experimental
characteristics and priors assumed in the previous case. We see that
this method gives generally good agreement with the nested sampling
computation, indicating that our calculations are robust. Some slight
differences are apparent, but this is expected as our version of the
Savage--Dickey method employs a Gaussian approximation for the
likelihood which may become poor at large distances from $\Lambda$CDM,
with the Fisher matrix method underestimating the covariance
matrix. For parameter estimation this is not a major concern, since
deviations from the Gaussian approximation occur in the tail of the
likelihood distribution, and quoted errors usually refer to the $68\%$
confidence intervals. However model selection calculations rely on
good modelling well into the tails of the distribution.

Having verified that our methods give similar results, henceforth we
will show results from the nested sampling method, since although it
is computationally more intensive it does not assume a Gaussian
likelihood.

\begin{figure*}
\subfigure
{\includegraphics[width=8cm]{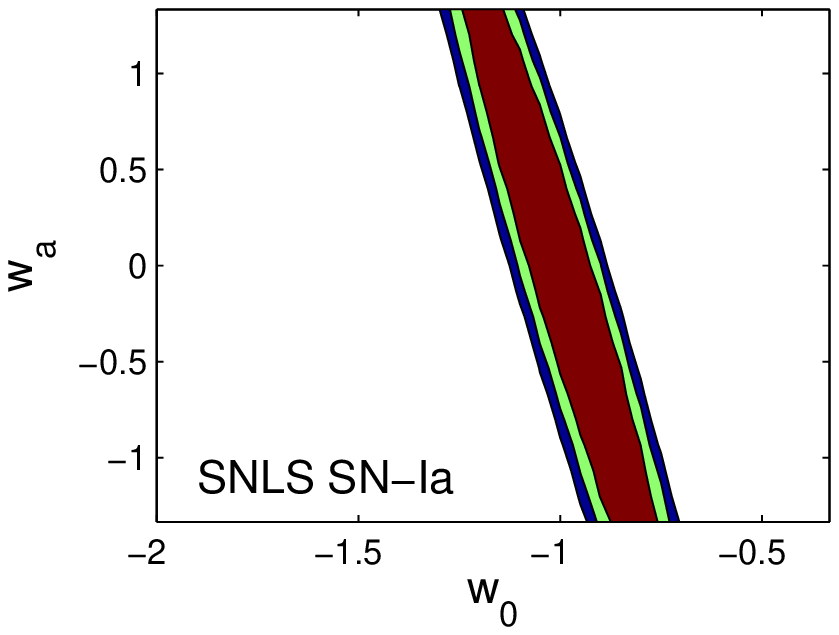}}
\subfigure
{\hspace*{0.5cm}
\includegraphics[width=8cm]{snap_pt01.eps}}\\ 
\subfigure
{\includegraphics[width=8cm]{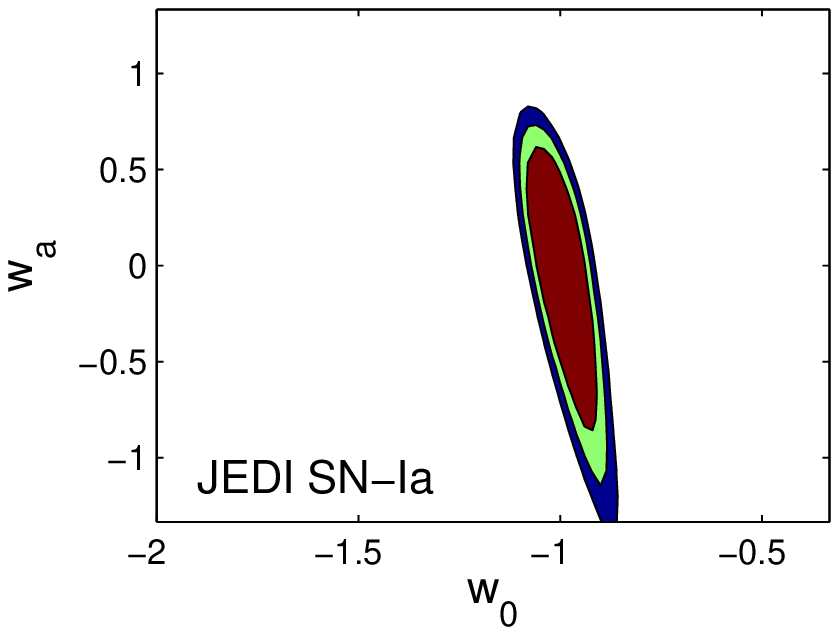}}
\subfigure
{\hspace*{0.5cm}
\includegraphics[width=8cm]{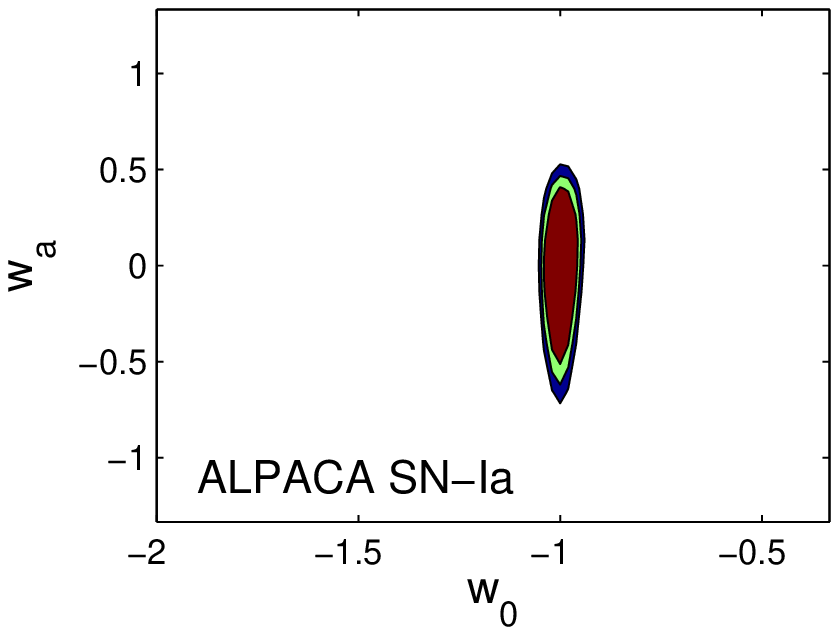}}\\
\subfigure
{\includegraphics[width=8cm]{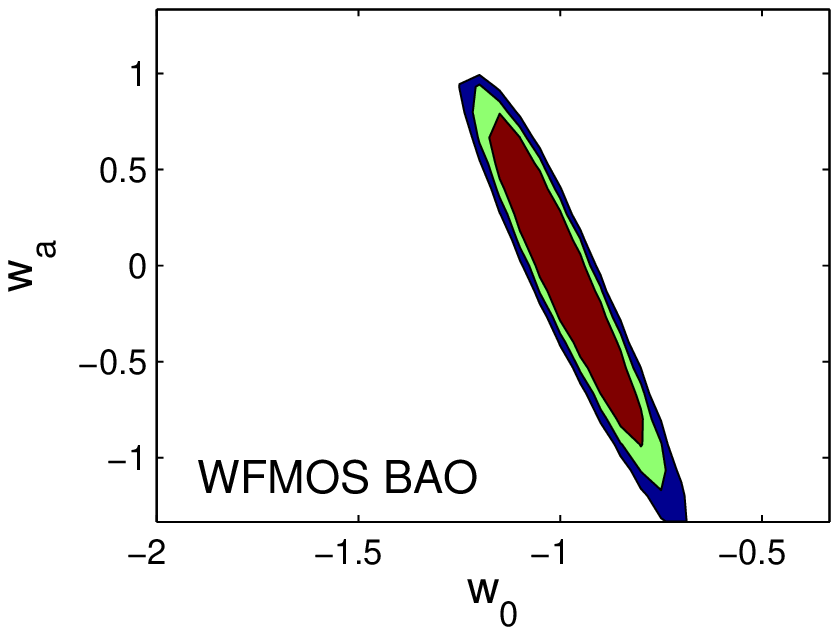}}
\subfigure
{\hspace*{0.5cm}
\includegraphics[width=8cm]{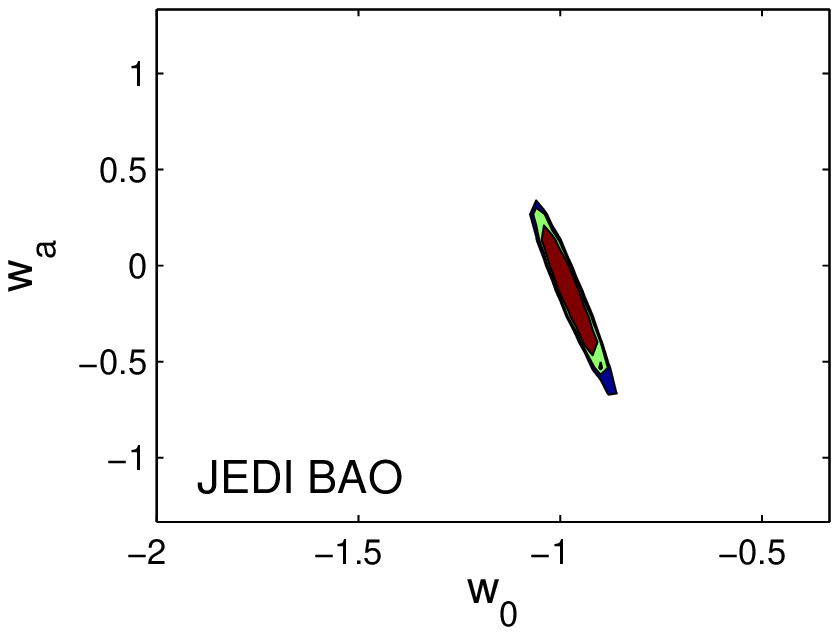}}\\
\caption{\label{fig:all} Bayes' factor forecasts for some future dark
energy surveys. Contours are shown for $\log(B_{01}) > 0$, $-2.5$ and
$-5$. An independent measurement of $\Omega_{{\rm m}}$ to $\pm 0.01$
is assumed. These plots show statistical uncertainties only, and
several of these experiments are likely to be dominated by
systematics.}
\end{figure*}

\subsection{Comparison with parameter error forecasting}

In this paper we are strongly advocating use of Bayes factor plots to
quantify experimental capabilities, for the reasons enumerated in
Section~\ref{ss:list}. It is useful to see explicitly what differences
this gives as compared to the traditional parameter forecast approach,
and so Figure~\ref{fig:SNAP2} shows a plot of likelihood contours,
obtained from a Markov chain Monte Carlo analysis of data simulated
for the SNAP supernova survey, using precisely the same assumptions as
Figure~\ref{fig:SNAP}, and assuming that $\Lambda$CDM is the true
model. 

We see they share the same general shape, and that the same principal
parameter degeneracy is picked out.  Obviously the two plots are
conceptually very different and so caution is needed in comparing.  We
see that the Bayes factor contours are significantly wider, indicating
that model selection sets a more stringent condition for dark energy
evolution to be supported by the data.  Indeed, the 95\% Fisher
parameter contour lies within the $\ln B_{01}=0$ contour where model
selection gives the models equal probability, hence by using the
Fisher matrix plot we could rule out $\Lambda$CDM with data that
actually favours it. It is fairly generic for that to be the case,
indicating that 95\% parameter estimation `results' tend not to be
robust under more sophisticated statistical analyses.  This is a
manifestation of Lindley's `paradox' as discussed by Trotta (2005) ---
that parameter values rejected under a frequentist test can
nevertheless be favoured by Bayesian model selection.

\subsection{Comparison of dark energy surveys}

We now turn to a comparison of the six dark energy surveys described
above. We stress once more that this comparison considers the
statistical uncertainties alone, and several of these experiments are
likely to be limited by systematics.  The criteria that enable the
systematics to be most effectively minimized are likely to be
different from those giving experiments raw statistical power. Further
we are working under the limitation of the particular $w_0$--$w_a$
parametrization; dark energy in reality could be different.

Figure~\ref{fig:all} shows the six surveys, the upper four being
supernova surveys and the lower two being the baryon acoustic
oscillation surveys. The innermost contoured region is where the
evidence of the $\Lambda$CDM model is greater than that of the
evolving dark energy model ($\ln B_{01} > 0$). The outer contours show
$\ln B_{01} =-2.5$ and $-5$ so that the data provides strong evidence
in favour of the evolving dark energy model.  As with parameter
estimation contours, the smaller the contours the more powerful the
experiment is.

As expected, we see a range of constraining powers depending on the
scale of the experiments. We also see that they broadly share the same
principal degeneracy direction, with slight rotations visible from the
different probing of redshift bins. The massive size of the expected
ALPACA dataset gives it the smallest contour area amongst SN-Ia
experiments, with its more limited redshift range making its
degeneracy more vertical.

The baryon oscillation probes share almost the same
principal degeneracy as the SNe-Ia; although they use the
angular-diameter distance rather than the luminosity distance these
two are related by the reciprocity relation and hence follow the same
degeneracy shape if the uncertainties in each redshift bin follow the
same shape. A probe which partly included the growth of structure,
such as weak lensing, would be expected to have a somewhat different
degeneracy; this has been shown using Fisher parameter contours for
the SNAP lensing survey though the rotation is still smaller than one
would like.

Note that the logarithms of the Bayes factors are additive, so if more
than one of these surveys happen, or if there are two independent
parts to a survey, then their Bayes factor plots can be added together
to give a net Bayes factor plot.

In addition to plotting Bayes factor contours, one can further
compress the information on how powerful an experiment is by computing
the area within a particular contour level, to give a single
`figure-of-merit'. Table~\ref{tab:areas} summarizes these areas, 
expressed in coordinate
units, for the six experiments, showing the area where $\ln B_{01}$
exceeds $-2.5$. Note that this corresponds to the parameter area in
which an experiment {\em cannot} strongly exclude $\Lambda$CDM, and
hence small numbers are better. For a more extensive discussion of
figures-of-merit for optimization of dark energy surveys, in a
parameter estimation rather than model selection framework, see
Bassett (2005) and Bassett, Parkinson \& Nichol (2005).

\begin{table}
\centering
\caption{Two experimental figures-of-merit: the areas in the
$\hat{w}_0$--$\hat{w}_a$ plane where $\ln B_{01}$ exceeds $-2.5$, and
the value of $\ln B_{01}$ at $\hat{w}_0=-1$ and $\hat{w}_a=0$. The
former measures the region of parameter space where the experiment
would not be able to exclude the $\Lambda$CDM model, while the latter
measures the strength with which the experiment would support
$\Lambda$CDM were it the true model. The $\ln B_{01}$ are additive
between surveys and for independent probes of dark energy within the
same survey.}
\begin{tabular}{lcc}
\hline
Experiment & Area & $\ln B_{01}(-1,0)$\\
\hline
SNLS     & 0.51 & 3.7 \\
SNAP     & 0.35 & 4.5 \\
JEDI SN  & 0.19 & 5.0 \\
ALPACA   & 0.08 & 6.1 \\
WFMOS    & 0.26 & 4.8 \\
JEDI BAO & 0.04 & 6.0 \\
\hline
\label{tab:areas}
\end{tabular}
\end{table}

\subsection{Support for $\Lambda$CDM}

We now consider the possibility of the experiments ruling out the dark
energy model in favour of $\Lambda$CDM, rather than the opposite which
we have focussed on thus far. Unlike parameter estimation methods,
Bayesian model selection can offer positive support in favour of the
simpler model. Because the simpler model is nested within the dark
energy model, it can never fit the data better, but it can benefit
from the volume effect of its smaller parameter space. All one needs
to do is read off the Bayes factor for the case where the fiducial
model is $\Lambda$CDM. Table~\ref{tab:areas} shows $\ln B_{01}$ at
$\hat{w}_0=-1$ and $\hat{w}_a=0$, i.e.~when $\Lambda$CDM is the true
model.

We find that this value is above 2.5 for all surveys, and above 5 for
several of them. Thus many of the surveys are capable of accumulating
strong evidence supporting $\Lambda$CDM over evolving dark
energy. This can be seen as another figure of merit quantifying the
power of experiments. Note again that the $\ln B_{01}$ are additive
between surveys and for independent probes of dark energy within the
same survey.

Note that the absolute value of this figure of merit is more sensitive
to the prior ranges chosen for the dark energy parameters, which set
the volume factor. However the relative comparison of surveys is again
not affected by this.

\begin{figure}
\centering
\includegraphics[width=8cm]{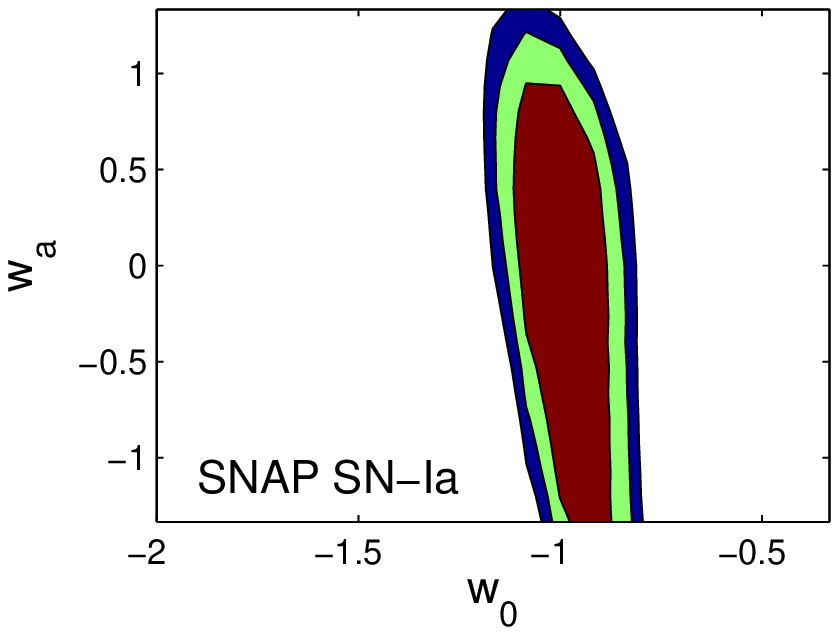}\\
\includegraphics[width=8cm]{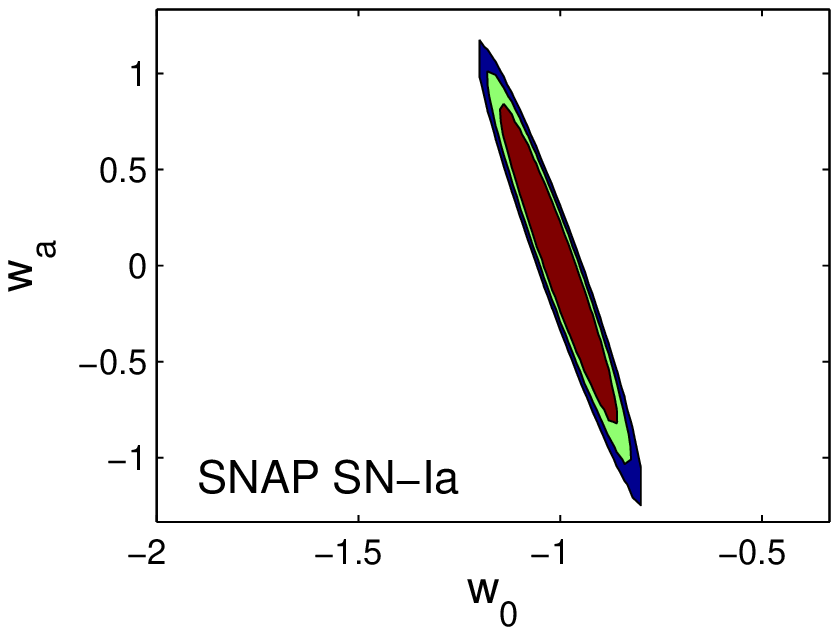}
\caption{\label{fig:3plots} Bayes' factor forecasts for SNAP assuming
different external knowledge of $\Omega_{\rm m}$.  Contours are again
shown for $\log(B_{01}) > 0$, $-2.5$ and $-5$. A Gaussian external
constraint on $\Omega_{\rm m}$ is assumed, of width 0.03 (top panel)
reflecting approximately the current level of uncertainty on it, and
0.003 (lower panel) reflecting an optimistic outcome.}
\end{figure}

\subsection{Variation of assumptions}

\label{s:variations}

We end by examining the effect of varying some of the assumptions that
went into the calculations, focussing on the SNAP supernova survey for
definiteness. We do this in three ways, one by changing the presumed
knowledge on $\Omega_{{\rm m}}$ that complements the dark energy
survey, one by looking at an alternative prior in the dark energy
model space, and finally by altering the assumed dispersion of
supernova luminosities and allowing for a simple model of systematics.

As mentioned before, the return on dark energy surveys is quite
sensitive to the availability of external constraints to remove
parameter degeneracies, particularly $\Omega_{{\rm m}}$ in the case of
the supernovae. Figure~\ref{fig:3plots} shows this effect for the SNAP
supernova survey, with different constraints on $\Omega_{{\rm m}}$ to
be compared with our standard assumption leading to the left panel of
Figure~\ref{fig:SNAP}. One sees a significant worsening of the Bayes
factor contours in the case of weaker knowledge on $\Omega_{{\rm m}}$.
 
Altering the constraint on $\Omega_{{\rm m}}$ can have a different
effect on different experiments. For instance, if it were more
stringent, then the difference between SNLS and SNAP or JEDI would be
greater --- the requirement for a more sensitive experiment would be
greater. Similarly the relative comparison is dependent on the nature
of dark energy itself; if a parametrization more complex than
$w_0$--$w_a$ proved necessary, it would be more important to make
precise high-redshift observations (e.g.~SNAP or JEDI versus ALPACA).

\begin{figure}
\centering
\includegraphics[width=8cm]{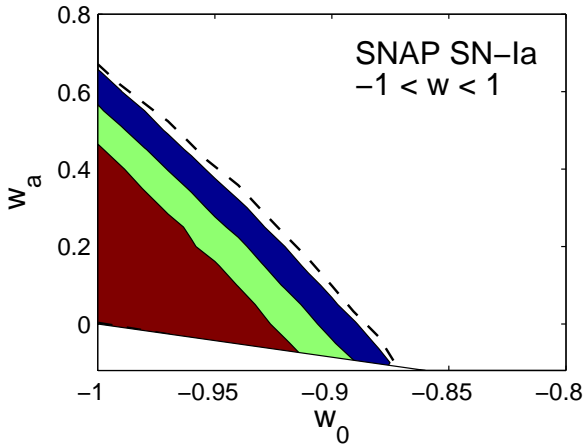}
\caption{\label{f:quint} SNAP Bayes' factor contours for the
quintessence prior on $w_0$ and $w_a$. The lower left-hand region is
cut off by the prior.  The dashed contour shows the location of the
$\ln B_{01} = -5$ contour for our full prior, as given in
Figure~\ref{fig:SNAP}.}
\end{figure}

Figure~\ref{f:quint} modifies our assumptions in a different way,
this time altering the prior on the dark energy parameters. It
assumes a prior appropriate to quintessence models, namely that
$w \ge -1$ at all redshifts. The evidence integral for the dark energy
model is then carried out over a narrower region in the dark energy
parameters, giving a boost to the evidence of the dark energy model
relative to $\Lambda$CDM. However the effect is small; the dashed
line shows where the outer contour lay with our full dark energy prior
and it has shrunk in only marginally.

Caldwell \& Linder (2005) classified quintessence models into freezing
and thawing models and delineated areas of the $w_0$--$w_a$ space
where those models typically lie. According to Figure~\ref{f:quint},
freezing models can only be decisively distinguished from $\Lambda$CDM
by the SNAP supernova survey if $\hat{w}_0 \ga -0.9$, and thawing
models if $\hat{w}_0 \ga -0.87$.

We have also investigated how changing the prior ranges on the dark
energy parameters alters the areas given in Table~\ref{tab:areas}. In
this case we narrowed the priors on $w_0$ and $w_a$ by a factor of two
in each direction. In cases where the posterior still lies within the
priors, this shifts the evidence by $\ln4 \simeq 1.4$ in favour of the
dark energy model. Unsurprisingly, we find this reduces the areas
within which $\Lambda$CDM cannot be excluded, typically by 10 to 20
per cent. Importantly, however, this change preserves the rankings of
the experiments.

\begin{figure}
\centering
\includegraphics[width=8cm]{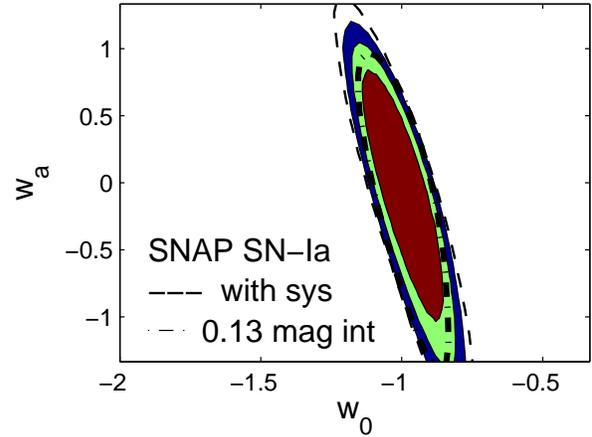}
\caption{\label{f:snapsys} The main contours match the left panel of
Figure~\ref{fig:SNAP}, showing the SNAP supernova
survey. Additionally, the dashed contour shows how the outer contour
shifts under a simple modelling of systematics, and the dot-dashed
contour shows how the outer contour would move if the magnitude error
were smaller.}
\end{figure}

Finally, in Figure~\ref{f:snapsys} we examine how the outer contour
would shift if a smaller magnitude dispersion were achieved (we take
$\delta_m = 0.13$), and separately under a standard (but crude)
modelling of possible systematics (see for example Kim et
al.~2004). The systematics have been modelled as an increased redshift
dependent uncertainty in magnitude of $(z/z_{\rm max}) \delta m_{\rm
sys}$ per redshift bin with $\delta m_{\rm sys} = 0.02$ mag, and added
in quadrature to the (intrinsic) statistical uncertainty. For SNAP,
this type of systematic has quite a small effect.  There can be other
types of systematics in the data, but we do not try to model them here
as the ability of different experiments to detect and (internally)
resolve systematics would be different and a proper study of
systematics and the required marginalization over them can only be
done once the data arrive.

\section{Conclusions}
\label{Conclusions}

In this paper we have introduced the Bayes factor plot as a tool for
assessing the power of upcoming experiments. It offers a full
implementation of Bayesian model selection as a forecasting tool. As
compared to the traditional parameter error forecasting technique, it
offers a number of advantages enumerated in
Section~\ref{ss:list}. Amongst those, perhaps the most important are
that observational data is simulated at each point in the plane,
rather than at a small number of fiducial models, and that the Bayes
factor plot properly captures the experimental motivation as being one
of model selection rather than parameter estimation.

As a specific example, we have used the Bayes factor plots to examine
a number of proposed dark energy surveys, concentrating on their
ability to distinguish between the $\Lambda$CDM model and a
two-parameter dark energy model. Figure~\ref{fig:all} indicates the
region of parameter space outside which the true model has to lie, in
order for the experiment to have sufficient statistical power to
exclude $\Lambda$CDM using model selection statistics.

An important caveat is that our plots do not show the effects of
systematics, which are likely to be the dominant uncertainty for many
of the experiments. This drawback is shared by parameter error
forecasts, and it is more or less the nature of systematic
uncertainties that they cannot be usefully modelled in advance of
actual observational data being obtained. In judging the true merit of
an experimental proposal, it is therefore essential to judge how well
structured it is for optimal removal of systematics, as well as
looking at its raw statistical power.

While we have focussed on dark energy as a specific application, the
concept of the Bayes factor plot has much broader applicability, and
is suitable for deployment in a wide range of cosmological contexts.

\section*{Acknowledgments}

PM, DP and ARL were supported by PPARC (UK), PSC by Columbia Academic
Quality Funds, and MK by the Swiss NSF. PM acknowledges Yun Wang for
helpful discussions regarding the application of the Fisher matrix
formalism to BAO observations. We thank Arlin Crotts for information
concerning JEDI, and Roberto Trotta for helpful discussions and
comments.

\appendix

\section{Marginalizing over simulated data}

In the main body of the paper, we have plotted the Bayes factor as a
function of the fiducial values of the dark energy parameters,
assuming particular values for the other parameters in the fiducial
model.  In general, however, the Bayes factor is a function of all the
fiducial model parameters, not just the ones of principal interest,
though a practical problem is that we cannot easily plot the evidence
ratio $B_{01}$ as a function of more than two variables
$\hat{\theta}$.

One solution is marginalization over the parameters that we are not
interested in, as one does in parameter estimation, assuming those
parameters to lie within the range motivated by present data.  As the
fiducial parameters $\hat{\theta}$ belong to the definition of the
data, we need to marginalize the evidences, $P(D|M)$, rather than the
Bayes factors. Formally, the marginalization must take place in data
space, so when we wish to integrate out a `nuisance' parameter
$\hat{\theta}_\mu$ which the data is a function of, we should take
into account a transformation factor $\sqrt{\sum_i (\partial
D_i/\partial\theta_\mu)^2}$, evaluated at each $\hat{\theta}$ along
the integral.  However, provided the evidence varies only weakly over
the relevant range of the fiducial models, or if our model depends
(nearly) linearly on its parameters, then this function is a constant
which cancels when computing the Bayes factor. In this case we can
just average the evidences. This will also conserve the relation
$B_{01}=1/B_{10}$.

In practice, the main determining factor in whether particular extra
parameters are justified by the data is the true values of those
parameters themselves, rather than values of the other
parameters. Often, then, one can choose fixed values of the
uninteresting parameters, presenting results on a slice through the
fiducial parameter space. Indeed, this turns out to be the case for
the dark energy surveys in this paper.

\section{Parameter error forecasting}\label{spf}

In this paper we are advocating the use of Bayes factors to quantify
the power of upcoming experiments, in place of parameter error
forecasts. For comparison, we provide a brief overview of parameter
error forecasting here, and discuss some of its features and
limitations. 

The idea is to simulate a sample of experimental data and then infer
the parameter uncertainties using standard likelihood analysis.  More
specifically, assuming a model $M$ specified by a set of parameters
$\theta=\{\theta_\mu\}$, a sample of data $D$ with the expected
experimental errors is generated for a particular fiducial model with
parameter values $\hat{\theta}$.  Then a likelihood $P(D|\theta,M)$ is
computed and the confidence intervals on the $\theta$ parameters are
inferred by computing a posterior parameter probability distribution
via Bayes' rule,
\begin{equation}
P(\theta|D,M) = \frac{P(D|\theta,M) P(\theta|M)}{P(D|M)} .
\end{equation}  
As a result the parameter uncertainties depend both on the experimental
characteristics and the choice of the fiducial model.

A simplified way of carrying out such an analysis is to use the Fisher
matrix approximation.  By construction the fiducial model parameter
values $\hat{\theta}$ maximize the likelihood. Hence expanding
$\ln{P(D|\theta,M)}$ to quadratic order in
$\delta{\theta}\equiv\theta-\hat{\theta}$ one obtains (Bond 1995;
Tegmark, Taylor \& Heavens 1997)
\begin{equation}
P(D|\theta,M)\sim 
\exp{\left(-\frac{1}{2}\sum_{\mu\nu}F_{\mu\nu}\delta{\theta_\mu}\delta
{\theta_\nu}\right)},\label{like}
\end{equation}  
which is a Gaussian approximation to the likelihood with zero mean and
with variance given by the inverse of the Fisher matrix $F_{\mu\nu}$,
where 
\begin{equation}
F_{\mu\nu}=\sum_{ij} C_{ij}^{-1}\frac{\partial D_i}{\partial\theta_\mu}
\frac{\partial D_j}{\partial\theta_\nu}\,.\label{fish}
\end{equation}
The sum is over all measurements and the partial derivatives are
evaluated at the fiducial model parameter values $\hat{\theta}_\mu$.
The matrix $C_{ij}$ is the data covariance matrix; for independent
measurements (e.g.~different supernovae) it simplifies to
$\sigma^2(D_i) \delta_{ij}$.  The parameter errors are then given by
the square root of the diagonal components of the covariance matrix,
$\sigma(\theta_\mu)=\sqrt{(F^{-1})_{\mu\mu}}$.  It is evident
from equation (\ref{fish}) that more accurate data, characterized by
smaller uncertainty $\sigma(D)$, provide larger Fisher matrix
components, hence smaller parameter errors. It can also be noticed
that for a given experiment the parameters which are better
constrained are those for which the partial derivatives are larger.

Since these derivatives are computed at the fiducial model, it is
natural to expect that the size of the projected errors varies for
different fiducial parameter values. These contours are usually
plotted with the aim of drawing a conclusion based on the true
model having different parameter values from those of the fiducial
model. But the dependence on the choice of fiducial model means that
there is no guarantee that the conclusions based on contours around
e.g.~the $\Lambda$CDM model can be used to rule that model out.

\begin{figure}
\centering
\includegraphics[width=8cm]{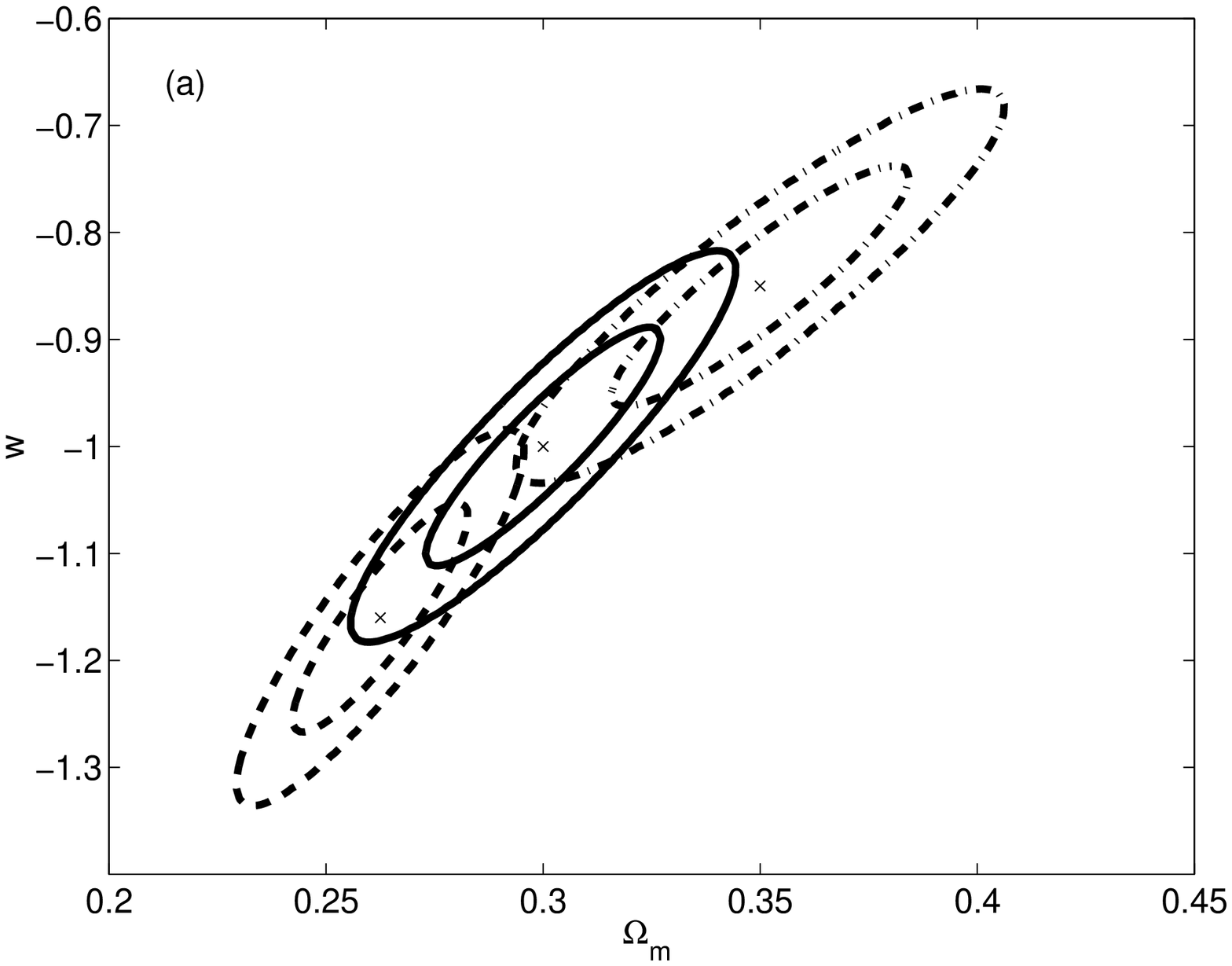}
\caption{The marginalized $68\%$ and $95\%$ confidence contours in
$\Omega_{{\rm m}}$--$w$ plane. The fiducial models are a $\Lambda$CDM
with $w=-1$ and $\Omega_{{\rm m}}=0.3$ (solid line), a dark energy
model with $w=-0.85$ and $\Omega_{{\rm m}}=0.35$ (dash-dot line), and
a phantom model with $w=1.16$ and $\Omega_{{\rm m}}=0.26$ (dash
line).}
\label{fig1}
\end{figure}

As an explicit example we compute the Fisher matrix errors of dark
energy parameters from SN-Ia luminosity--distance measurements.  We
assume experimental characteristics from the proposed SNAP mission as
discussed in Kim et al.~(2004).  We consider two different dark energy
models, one parametrized by a constant equation of state parameter
$w$, and a second by the two-parameter equation of state family of
equation (\ref{lp}).  We assume an independent measurement of
$\Omega_{{\rm m}}$ with uncertainty $\pm 0.03$ to reduce parameter
degeneracies, and compute the marginalized confidence contours around
different fiducial models in $\Omega_{{\rm m}}$--$w$ and $w_0$--$w_a$
planes respectively.

In Figure~\ref{fig1} we plot the $68\%$ and $95\%$ ellipses around
three models: a $\Lambda$CDM model with $w=-1$ and $\Omega_{{\rm
m}}=0.3$, a dark energy model with $w=-0.88$ and $\Omega_{{\rm
m}}=0.35$, and a phantom model with $w=-1.16$ and $\Omega_{{\rm
m}}=0.26$. The alignment of the strongest degeneracy line differs
amongst the models. This is because the degeneracy in the
$w-\Omega_{{\rm m}}$ plane is not a straight line, but rather a curve
(see for instance Weller \& Albrecht 2001).  Notice also that the
ellipses around the fiducial models have different sizes.  As
Fig.~\ref{fig1} shows, if the true model lies on say the 95\%
confidence limit of the $\Lambda$CDM data, one cannot necessarily
presume that the $\Lambda$CDM model would lie on the 95\% confidence
limit of data simulated for the true model. It is possible to compute
a contour indicating the locus of the fiducial models for which
$\Lambda$CDM lies at their 95\% confidence limit, and indeed such a
locus is shown in Figure 1 of Kratochvil et al.~(2004), but
constructing it is a rather cumbersome procedure.

\begin{figure}
\centering
\includegraphics[width=7.8cm]{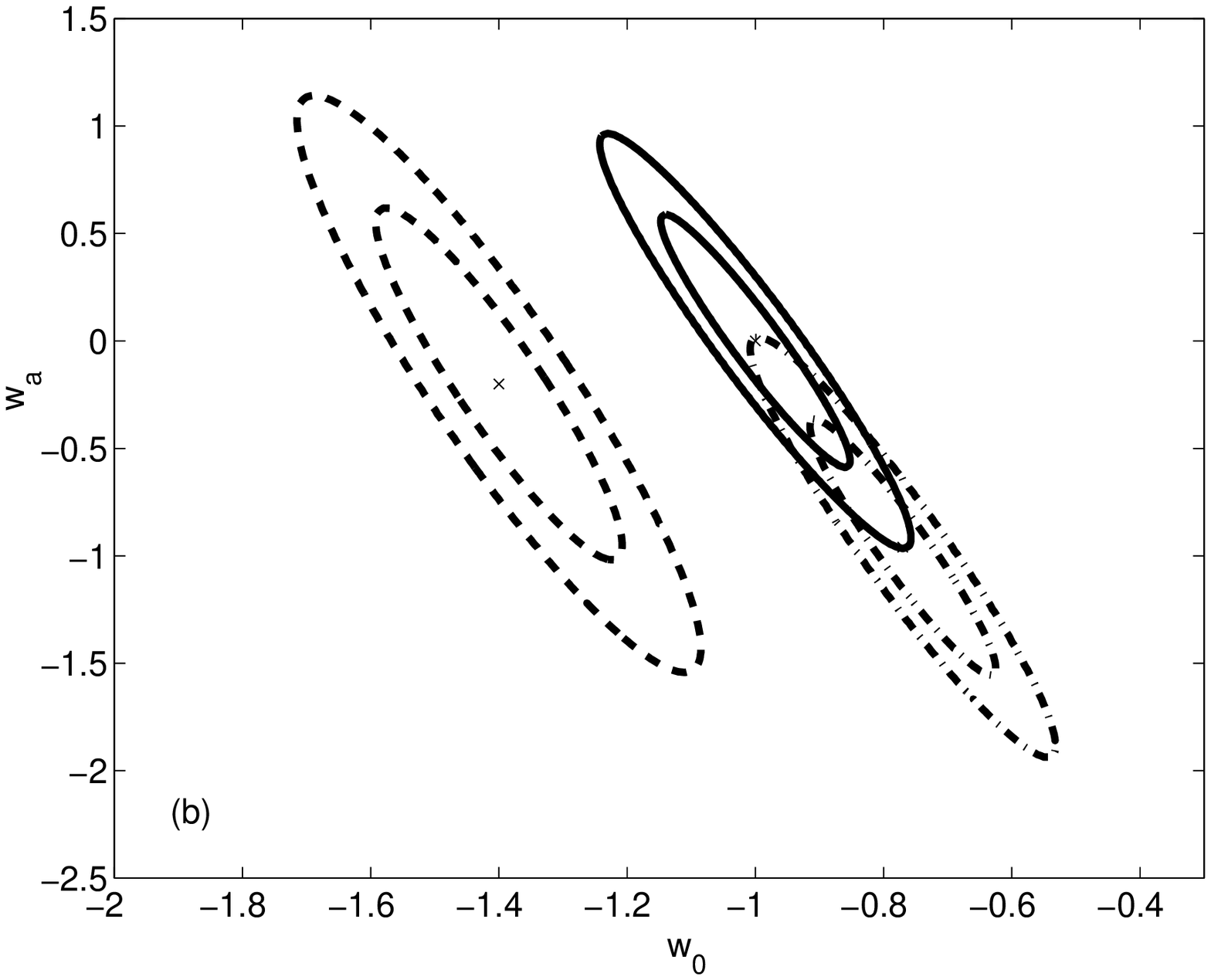}
\caption{Marginalized $68\%$ and $95\%$ contours in $w_0$--$w_a$
plane.  The fiducial models are $\Lambda$CDM (solid line), a dark
energy model with $w_0=-0.8$ and $w_a=-1$ (dash-dot line), and a
phantom model with $w_0=-1.4$ and $w_a=-0.2$ (dash line). For all
three models $\Omega_{{\rm m}}=0.3$.}
\label{fig2}
\end{figure}

This drawback turns out to be less severe for dark energy models
parametrized by equation~(\ref{lp}). In Figure~\ref{fig2} we plot the
$68\%$ and $95\%$ ellipses in the $w_0$--$w_a$ plane with
$\Omega_{{\rm m}}=0.3$ around a $\Lambda$CDM model, a phantom model
with $w_0=-0.8$ and $w_a=-1$ lying along the degeneracy line of the
$\Lambda$CDM, and a constant phantom model with $w_0=-1.4$ and
$w_a=-0.2$.  The dependence on the fiducial model is still present,
since the ellipses become larger as the fiducial model shifts
orthogonal to the principal degeneracy direction towards more negative
equation of state values. Fiducial models along the same degeneracy
line whose $95\%$ contours include the $\Lambda$CDM model are within
the $95\%$ ellipse of the $\Lambda$CDM as well.

\bsp

\begin{thebibliography}{}

\bibitem[Aldering et al.~(2004)]{SNAP} Aldering G. et al.~(SNAP
	collaboration), 2004, astro-ph/0405232
\bibitem[Astier et al.~(2006)]{SNLS2} Astier P. et al.~(SNLS
	collaboration), 2006, A\&A, 447, 31
\bibitem[Bassett (2005)]{Bassett:2004st} Bassett B. A., 2005,
	Phys. Rev. D, 71, 083517
\bibitem[Bassett et al.~(2004)]{Bassett04}
	Bassett B. A., Corasaniti P. S., Kunz M., 2004, ApJ, 617, L1
\bibitem[Bassett et al.~(2005)]{Bassett:2004np} Bassett B. A., Parkinson D.,
	Nichol R. C., 2005, ApJL, 626, L1
\bibitem[Bond (1995)]{Bond95} Bond J. R., 1995, Phys. Rev. Lett., 74, 4369
\bibitem[Bond, Efstathiou \& Tegmark (1997)]{Bond97}
        Bond J. R., Efstathiou G., Tegmark M., 1997, MNRAS, 291, L33
\bibitem[Caldwell and Linder (2005)]{CL05} Caldwell R. R., Linder E. V.,
	2005, Phys. Rev. Lett, 95, 141301
\bibitem[Chevallier \& Polarski (2001)]{CP01} Chevallier M., Polarski
	D., 2001, Int. J. Mod. Phys., D10, 213
\bibitem[Corasaniti \& Copeland (2003)]{Coras02} 
        Corasaniti P. S., Copeland E. J., 2003, Phys. Rev. D, 67, 063521
\bibitem[Corasaniti et al.~(2005)]{CLCB} Corasaniti P. S., LoVerde M.,
	Crotts A., Blake C., 2005, astro-ph/0511632
\bibitem[Crotts et al.~(2005)]{JEDI} Crotts A. et al.~(JEDI
        collaboration), 2005, astro-ph/0507043
\bibitem[Dickey (1971)]{D71} Dickey J. M., 1971, Ann. Math. Stat., 42,
	204 
\bibitem[Efstathiou \& Bond 1999]{Efs98} Efstathiou G., Bond J. R., 1999,
	MNRAS, 304, 75 
\bibitem[Gregory (2005)]{Gregory} Gregory P., 2005, {\it Bayesian Logical 
	Data Analysis for the Physical Sciences}, Cambridge University
	Press
\bibitem[Jaffe (1996)]{J96} Jaffe A., 1996, ApJ, 471, 24
\bibitem[Jeffreys (1961)]{Jeffreys} Jeffreys H., 1961, {\it Theory of
	Probability}, 3rd edition, Oxford University Press 
\bibitem[Jungman et al.96]{Jetal} Jungman J., Kamionkowski M., Kosowsky A., 
	Spergel D. N., 1996, Phys. Rev. D, 54, 1332
\bibitem[Kass \& Raftery 1995]{KR95} Kass R. E., Raftery A. E., 1995, J.
	Amer. Stat. Assoc., 90, 773
\bibitem[Kim et al. 2004]{snap} Kim A. G., Linder E. V., Miquel R.,
	Mostek N., 2004, MNRAS, 347, 909
\bibitem[Knox(1995)]{Knox95} Knox L., 1995, Phys. Rev. D, 52, 4307
\bibitem[Kratochvil et al.(2004)]{Kratochvil04} Kratochvil J., Linde A.,
	Linder E. V., Shmakova M., 2004, JCAP, 0407, 1
\bibitem[Linder03]{Linder03} Linder E. V., 2003, Phys. Rev. Lett., 90,
	091301 
\bibitem[LinHut05]{LinHut} Linder E.V., Huterer D., 2005, Phys. Rev. D,
	 72, 043509
\bibitem[MacKay(2003)]{MacKay03} 
	MacKay D. J. C., 2003, {\it Information theory, inference and
	learning algorithms}, Cambridge University Press
\bibitem[Marshall, Hobson \& Slosar(2003)]{Marshall03}
	Marshall P. J., Hobson M. P., Slosar A., 2003, MNRAS, 346, 489
\bibitem[Mukherjee et al.~(2006)]{MPL} Mukherjee P., Parkinson D.,
	Liddle A. R., 2006, ApJL, 638, L51
\bibitem[Pogosian et al. 2005]{pogosian05} Pogosian L., Corasaniti
	P. S., Stephan-Otto C., Crittenden R., Nichol R., 2005,
	Phys. Rev. D, 72, 103519
\bibitem[Pritchet et al.(2004)]{SNLS} Pritchet C. J.~(SNLS
	Collaboration), 2004, astro-ph/0406242
\bibitem[Saini, Weller \& Bridle(2004)]{Saini04} 
	Saini T. D., Weller J., Bridle, S. L., 2004, MNRAS, 348, 603
\bibitem[Seo \& Eisenstein 2003]{SE03}
        Seo H.-J., Eisenstein D. J., 2003, ApJ, 598, 720
\bibitem[Skilling(2004)]{Skilling04} 
	Skilling J., 2004, in {\em Bayesian Inference and Maximum Entropy 
	Methods in Science and Engineering}, ed. R. Fischer et al.,
	Amer. Inst. Phys., conf. proc., 735, 395 (available at {\tt 
	http://www.inference.phy.cam.ac.uk/bayesys/}).
\bibitem[Tegmark et al. 2000]{teg00} Tegmark M., Eisenstein D. J., Hu
	W., de Oliveira-Costa A., 2000, ApJ, 530, 133
\bibitem[Tegmark et al. 1997]{teg97} Tegmark M., Taylor A., Heavens
	A., 1997, MNRAS, 480, 22
\bibitem[Trotta(2005)]{Trotta05} Trotta R., 2005, astro-ph/0504022.
\bibitem[Verdinelli \& Wasserman (1995)]{VW} Verdinelli I, Wasserman
	L., 1995, J. Amer. Stat. Assoc., 90, 614
\bibitem[Wang \& Tegmark (2004)]{WT} Wang Y., Tegmark M., 2004,
	Phys. Rev. Lett., 92, 241302
\bibitem[Weller \& Albrecht (2001)]{WA} Weller J., Albrecht A., 2001, Phys.
	Rev. Lett., 86, 1939
\bibitem[Zal97]{Zal97} Zaldarriaga M., Spergel D. N., Seljak U., 1997,  
	ApJ, 488, 1
\end{thebibliography}
\end{document}